# Vanadium Dioxide: Metal-Insulator Transition, Electrical Switching and Oscillations. A Review of State of the Art and Recent Progress

Alexander Pergament[1,2,*)], Aurelian Crunteanu[1], Arnaud Beaumont[1],

Genrikh Stefanovich[2], and Andrey Velichko[2]

[1] *XLIM Research Institute, UMR7252 CNRS/ Université de Limoges, LIMOGES Cedex FRANCE*

[2]*Department of Physical Engineering, Petrozavodsk State University, PETROZAVODSK Republic of Karelia RUSSIA*

**Abstract.** Vanadium dioxide is currently considered as one of the most promising materials for oxide electronics. Both planar and sandwich thin-film metal/oxide/metal devices based on $VO_2$ exhibit electrical switching with an S-shaped *I-V* characteristic, and this switching effect is associated with the metal-insulator transition (MIT). In an electrical circuit containing such a switching device, relaxation oscillations are observed if the load line intersects the *I–V* curve at a unique point in the negative differential resistance region. All these effects are potentially prospective for designing various devices of oxide electronics, such as, particularly, elements of dynamical neuron networks based on coupled oscillators. In this paper, we present a brief review of the state of the art and recent advances in this though only freshly emerged, but exciting and rapidly developing field of research. We start by describing the experimental facts and theoretical models, mainly on the basis of the Mott MIT in vanadium dioxide as a model object, of the switching effect with especial emphasis on their applied potentialities for oxide electronics, and then turn to our recent results concerning synchronization of an array of coupled oscillators based on $VO_2$ switches. In conclusion, we try to formulate an outlook on how these findings can contribute to the future development of bio-inspired neuromorphic devices implementing the associative memory function, that is, non-Boolean computing with artificial neural networks[**)].

## 1. Introduction

The unprecedented diversity of physical properties exhibited by transition metal oxides offers immense prospects for various electronic applications. The term "oxide electronics" have emerged not so long ago in the everyday-life of scientific literature, but already firmly taken its place [1]. The point is that the modern IT revolution is based on technological progress which enables an exponentially growing enhancement of the performance of electronic devices. During all the history of the development of electronic components, from a vacuum diode to modern highly integrated ICs with nanometer scale of individual elements, the question of the physical limitations on the further progress in this area arose repeatedly.

---

After the invention of an IC by J. Kilby and R. Noyce in 1958 [2], the amount of devices is continuously increased from 1960s by means of device miniaturisation, following the so-called Moore's law. In recent years, the issue of constraints for standard Si-based electronics has been widely discussed in the scientific literature, which is primarily associated with the possibility of further scaling toward nano-size. In this regard, in the 2007 edition of The International Technology Roadmap for Semiconductors (ITRS, http://www.itrs.net), a new section has appeared, namely "Emergent Research Device Materials", which indicates the need to develop a new generation of devices based on new physical principles [3].

Alternative approaches are based on another mechanism (as compared to the field effect in Si CMOS FETs) or even on a drastic change in computational paradigm or architecture (quantum computers, neuroprocessors). Amongst the approaches utilizing new physical mechanisms, one can list, for example, spintronics, superconducting electronics, single-electronics, molecular electronics, soletronics (single atom electronics) [1]. One of such novel directions, oxide electronics, is based on the idea of application of unique properties and physical phenomena in strongly correlated transition metal oxides (TMO). Metal-insulator transition (MIT) [4] belongs to the class of the aforementioned phenomena, and many TMOs, e.g. vanadium dioxide, undergo MITs as functions of temperature or electric field [4, 5].

Vanadium dioxide is currently considered as one of the most promising materials for oxide electronics. Both planar and sandwich thin-film MOM devices based on $VO_2$ exhibit electrical switching with an S-shaped $I$-$V$ characteristic, and this switching effect is associated with the MIT. In an electrical circuit containing such a switching device, relaxation oscillations are observed if the load line intersects the $I$–$V$ curve at a unique point in the negative differential resistance (NDR) region. All these effects are potentially prospective for designing various devices of oxide electronics, such as, particularly, the Mott-FET (filed effect transistor based on the Mott MIT material) [5] or elements of dynamical neuron networks based on coupled oscillators [6].

In this brief review we discuss the state of the art and recent advances in this though only freshly emerged, but exciting and rapidly developing field of research. We start by describing the experimental facts and theoretical models, mainly on the basis of the Mott transition in vanadium dioxide as a model object, of the switching effect with especial emphasis on their applied potentialities for oxide electronics, and then turn to our recent results concerning synchronization of an array of coupled oscillators based on $VO_2$ switches. The latter findings are of crucial importance for future development of bio-inspired neuromorphic devices implementing the associative memory function, that is, non-Boolean computing with artificial neural networks [6, 7].

## 2. MIT and electronic switching in VO₂

### 2.1. Overview of experimental data

Since the discovery of the MIT in $VO_2$ [8], there has been a lot of interest in this kind of phase transition, as it is accompanied by a large change in the electrical resistivity and optical properties. This transition occurs at $T_t = 340$ K (Fig.1). In single crystals, the resistivity change reaches a factor of $10^5$ over a temperature range of 0.1 K [9]. Hysteresis associated with this transition is of about 2 K. The conductivity jump and the narrowness of the hysteresis loop is a very good indication of how close the stoichiometry is to $VO_2$. Small deviations destroy the sharpness of the transition and increase the hysteresis width. The crystalline state of the material has an influence too: polycrystalline material will have a broader transition than single crystals. The transition temperature also depends on the crystalline state and oxygen non-stoichiometry. As a rule, the MIT in $VO_2$ is to a certain degree suppressed in thin films as compared to single crystals.

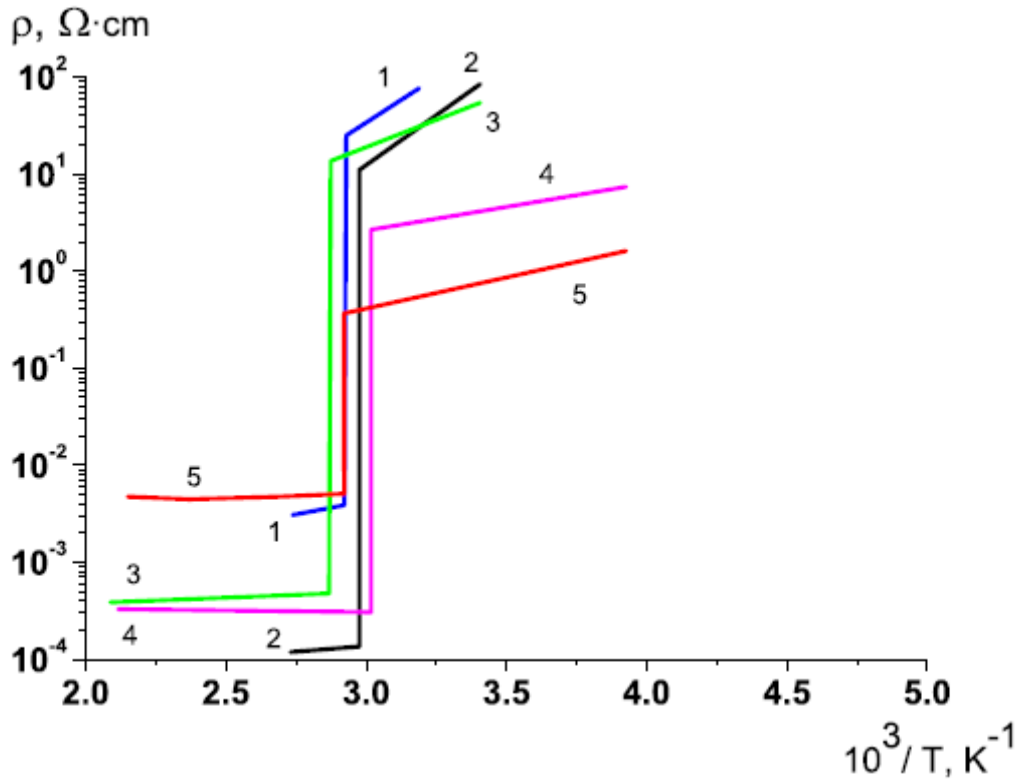

**Fig. 1.** Resistivity of $VO_2$ single crystals along the *c* axis as a function of reciprocal temperature (1 - 5: data from different works for samples of various degrees of perfection; for references see [4], 1st edition of 1974) [5].

At high temperatures, vanadium dioxide has a rutile-type structure with vanadium atoms equally spaced along the corresponding $c_r$-axis. At the transition to the low-temperature state, these vanadium atoms pair along the $c_r$-axis with a slight twist which leads to a monoclinic symmetry. The gap in the insulating low-temperature phase is about 0.7-1.0 eV (depending on stoichiometry and other factors).

Despite the many efforts made towards obtaining an understanding of their electronic behavior, the vanadium oxides still pose many open questions. This fully applies to $VO_2$ as far as the nature of transition is concerned. Some authors related its phase transition to a Mott-Hubbard scenario, whereas others attributed it to electron-phonon coupling (a Peierls mechanism) on the basis of the crystal symmetry change. In the recent monograph [10], it is stated that the metal-insulator transition in $VO_2$ has a combined nature, having the features of both Mott and Peierls transitions.

Nonetheless, many experimental facts indicate that the transition might be initiated by an increase in the free charge carrier density (without heating to $T = T_t$ and without affecting the lattice, i.e. not under, for example, doping or pressure) under photo-generation [11], injection [12], or high-field generation at switching [13]. These results evidence that the electron-electron interaction is of importance for a correct description of the transition. This is supported by studies of the MIT in thin films of amorphous vanadium dioxide obtained by anodic oxidation [14, 15].

It should be emphasized however that the question about the MIT mechanism in $VO_2$, posed as "either Mott or Peierls transition" [16], has no sense because, actually, this mechanism is essentially dual [17]. Meanwhile, one should be aware of the fact (and take this fact into account when analyzing the capabilities of $VO_2$-based electronic devices) that the initiating mechanism is still the correlation-driven electronic Mott transition [5].

In two-terminal MOM devices with vanadium dioxide, electrical switching due to MIT is observed [18-27] (Fig. 2). This switching effect is described in terms of the current-induced Joule heating of the sample up to $T = T_t$, which has been confirmed by the direct IR-radiation measurements of the switching channel temperature [20, 27]. In these early works, switching has been observed in single crystals [21], planar thin-film devices [19, 20, 27], as well as in vanadate glasses [22, 23], $VO_2$-containing ceramics [26], $V_2O_5$-gel films [24, 25] and anodic oxide films on vanadium [18]. When as-prepared samples do not consist of pure vanadium dioxide, preliminary electroforming is required, resulting in the formation of the $VO_2$ containing channel [18, 22, 24]. For such samples the threshold voltage decreases with temperature and reaches zero at $T \sim T_t$.

In this case, the switching mechanism is well described by a simple thermal model often termed as the "model of critical temperature" [19], provided that the ambient temperature $T$ is not much lower than $T_t$:

$$V_{th} \sim (T - T_t)^{1/2}, \qquad (1)$$

that is, the squared threshold voltage linearly tends to zero at $T \rightarrow T_t$, see Fig. 2(c).

However, at a sufficiently low temperature, and in sandwich thin-film nanostructures (that is, in high electric fields), the possible influence of electronic effects on the MIT should be taken into account. A field-induced increase in charge carrier density will act to screen Coulomb interactions, leading to the elimination of the Mott-Hubbard energy gap at $T < T_t$ [5, 12, 13]. This non-thermal (electronic) mechanism of switching in $VO_2$ is considered in the next section.

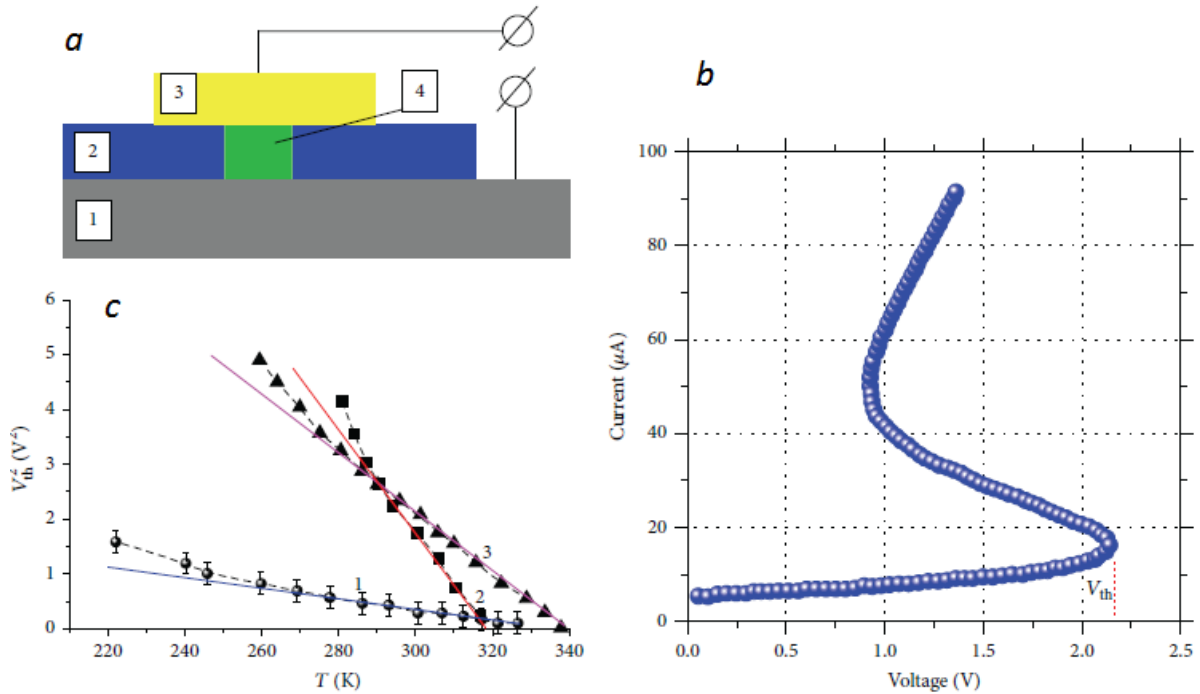

**Fig.2.** (a) The sandwich switching structure (schematic) based on anodic vanadium oxide (AVO) [5]: 1 – vanadium metal substrate, 2 – anodic oxide film, 3 – Au electrical contact, 4 – $VO_2$ switching channel; (b) the *I-V* characteristic after electroforming for one of the samples at room temperature. (c) Squared threshold voltage as a function of temperature for vanadium oxide-based switch: 1, 2, 3 – three different samples.

Finally, it should be noted that although the switching effect in vanadium dioxide has long been known, recently this problem has received considerable attention once again, primarily due to the promising potential applications in emerging oxide electronics (what these applications are, we have briefly mentioned in the introduction, and will discuss in more detail below in Section 3). In this regard, a large and ever-growing number of works dedicated to switching in vanadium dioxide have being published at present. Without a possibility to analyze or even mention all of them here, we refer to the recent reviews [5, 28, 29] where the reader can find corresponding discussions.

## 2.2. Electronically-driven switching in $VO_2$

Application of an electric field to a Mott insulator, taking also into account the current flow, will result in both the thermal generation of additional carriers due to the Joule heat, and in the field-induced generation. The latter is due to the autoionization caused by the Coulomb barrier lowering – a phenomenon analogous to the Poole-Frenkel effect [13], when the density depends on the electrical field strength $E$ as follows:

$$n = N_0 \exp\left(-\frac{W - \beta\sqrt{E}}{kT}\right), \qquad (2)$$

where $N_0$ is a constant independent of the field and only slightly dependent on temperature, $W$ – the conductivity activation energy, and $\beta = (e^3/\pi\varepsilon_\infty\varepsilon_o)^{1/2}$ – the Poole-Frenkel constant; $\varepsilon_\infty$ is the high-frequency dielectric permittivity.

When the total concentration of free carriers reaches the value of $n = n_c$, the transition into the metal state, i.e. switching of the structure into the ON-state, occurs. In a material with the temperature-induced MIT, like $VO_2$, this switching (in a relatively low field) occurs just due to the heating of the switching channel up to $T = T_t$ according to the critical temperature model, equation (1). The above mentioned value of $n_c$ is determined by the Mott criterion:

$$a_H n_c^{1/3} \approx 0.25, \qquad (3)$$

where $a_H = \varepsilon\hbar^2/m^*e^2$ is the effective Bohr radius, $m^*$ – the effective mass of charge carriers and $\varepsilon$ – the static dielectric permittivity.

In the paper [13], it has been reported on the switching mechanism in vanadium dioxide. This mechanism is shown to be based on the electronically-induced Mott insulator-to-metal transition occurring in conditions of the non-equilibrium carrier density excess in the applied electric field. The model has been developed on the basis of the measurements of *I-V* characteristics of M/AVO/M sandwich structures (Fig. 3). It has been shown that, as the electric field strength increases, the transition temperature (i.e. the temperature at which the switching event inside the channel occurs) decreases. Numerical simulation of the free charge carrier density has been made (Fig. 4, curve 1) using the relation $\sigma = ne\mu$, where $\sigma$ is the channel conductivity obtained from the data of Fig. 3, and $\mu = 0.5$ cm$^2$ V$^{-1}$ s$^{-1}$ is the mobility of electrons in vanadium dioxide. One can see that at $T_0 < 200$ K ($E > 10^5$ V cm$^{-1}$), for $\mu = $ const, the value of $n$ becomes less than the critical Mott density $n_c$ in equation (3), which for vanadium dioxide has been estimated to be $n_c \sim 3 \cdot 10^{18}$ cm$^{-3}$ (dotted line in Fig. 4).

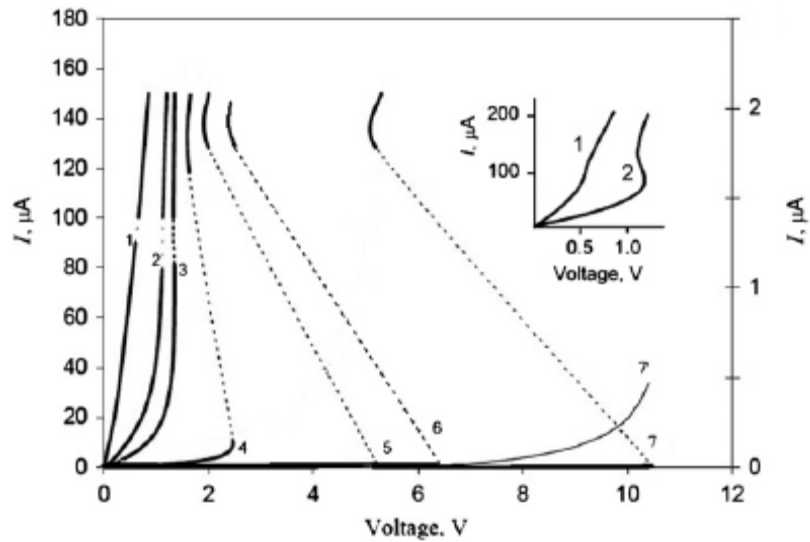

**Fig. 3**. *I-V* curves at different temperatures in the range 15 – 300 K [13]. $T_0$: (1) 293 K, (2) 241 K, (3) 211 K, (4) 144 K, (5) 91 K, (6) 70 K, (7) 15 K. Curve (7') represents the OFF state for $T_0 = 15$ K (right-hand current axis), and the inset shows curves (1) and (2) on a large scale.

If however we take into account the temperature dependence of the mobility, this discrepancy can be eliminated [30]. It has been shown [31] that the conductivity of a VO$_2$ switching channel can be written as

$$\sigma \sim \frac{1}{T^{3/2}} \exp(-E_a / k_B T + k_B T / \xi), \tag{4}$$

where $E_a$ is the conductivity activation energy, and $\xi$ is a constant of the small polaron hopping conduction theory which takes into account the influence of thermal lattice vibrations onto the resonance integral [31]. In equation (4), the first term in the exponent is responsible for the band conduction, whereas the second one characterizes the temperature dependence of the hopping mobility. Therefore the data on $n_m(T)$ (Fig. 4, curve 1) should be corrected by means of division by a factor corresponding to the mobility temperature dependence, which could be obtained from Eq.(4) [30]:

$$\mu(T) \sim (1/T) exp(k_B T / \xi). \tag{5}$$

These corrected values of $n_m$ are presented by curve 2 in Fig. 4. One can see that, within an order of magnitude, these values correspond to the Mott critical concentration.

Thus, the switching mechanism is based on the electronically-induced Mott metal-insulator transition taking into consideration the dependence of the carrier density on electric field, equation (2), and the temperature dependence of the electron mobility in VO$_2$, equation (5).

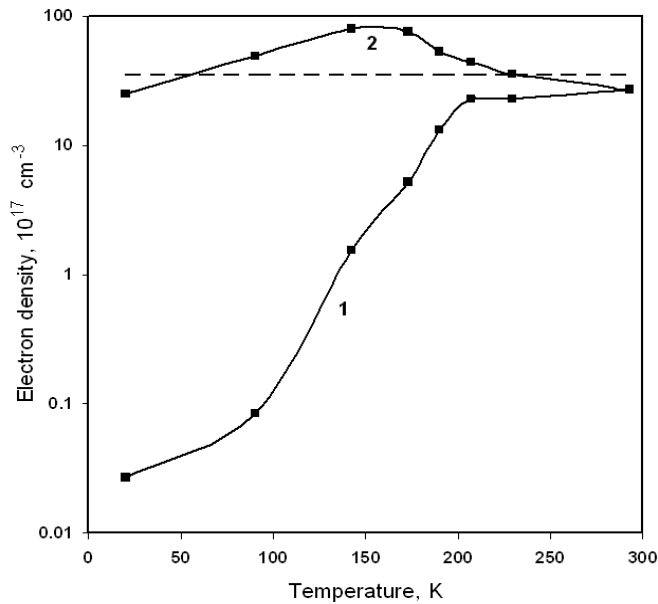

**Fig. 4.** Maximum electron density in the switching channel as a function of ambient temperature; in fact, $n_m$ depends on the threshold electric field $E_{th}$, and the latter, in turn, depends on $T_0$. The calculations are made for the cases of $\mu = $ const (curve 1) and $\mu \sim T_m^{-1} exp(k_B T_m / \xi)$ (curve 2). Here $T_m$ ($\neq T_0$) is the temperature in the channel centre calculated on the basis of mathematical simulating [13].

### 2.3. Relaxation oscillations in circuits with VO$_2$ switches

In a scheme with an S-NDR device, at a high enough d.c. bias ($V_0$), relaxation oscillations are observed if the load line intersects with the *I–V* curve at a unique point in the NDR region (Fig. 5) [5]. The frequency of this oscillation ($f_r = 1/T_r$) depends on the threshold parameters and, consequently, it depends on external parameters, e.g. on temperature and pressure. Sensors with the frequency output have a number of evident advantages: high sensitivity and stability, convenience of processing of a signal, capability of remote measurements, low noise susceptibility.

The expression for $T_r$ one can obtain from the known exponential voltage *vs*. time dependence for the charging of a capacitor:

$$V(t) = V_0[1 - \exp(-t/\tau)], \tag{6}$$

whence:

$$T_r = \tau \ln \frac{V_0 - xV'}{V_0 - xV_{th}}, \tag{7}$$

where $x = (R_L/R_0)+1$, $\tau = R_L C/x$ and $R_0$ is the OFF-state resistance of the switch [32]. In the absence of a parallel-connected capacitor, $C$ is the switch self-capacitance determined by its geometrical dimensions and the value of $\varepsilon(f, T_0)$ of VO$_2$. Fig. 5(d) shows the experimental temperature dependencies of the oscillation frequency for three M/AVO/M samples.

Much attention in the literature is also paid to the noise and chaotic dynamics at switching in vanadium dioxide, primarily because of their importance in bolometric applications [33-35]. Particularly, generation of stochastic oscillations has been observed in [36]. For sufficiently large amplitudes of the periodic signal, a new nonlinear effect in stochastic resonance has been found. In these experiments, a sin-wave generator was connected in series with the DC source. In the presence of a weak internal noise, it is possible to observe the phenomenon of locking the mean switching frequency, see Fig. 6. At first, the frequency of switching coincides with the signal frequency $\Omega$. Then, as the $\Omega$ increases over the natural system frequency $f_r$, the switching period becomes twice as greater in comparison with $T_L=1/\Omega$ (Fig. 6 (c)). Between these two states, a chaotic regime appears with the phase diagram of *I* and *V* shown in Fig. 6(b). Thus, there exist two limit cycles, between which the phase trajectory moves chaotically and fills in the space between them. As the frequency continues to increase, the described process is repeated, i.e. the oscillations with $T = NT_L$ ($N = 3, 4, 5…$) appear. The same transition between different oscillating behaviors is observed when varying the amplitude of the sin-wave signal or the DC voltage $V_0$, not only the frequency $\Omega$. The above-described chaotic behavior represents the intermittent chaos [36], i.e. the random transitions between two limit cycles (attractors) during the period-adding process.

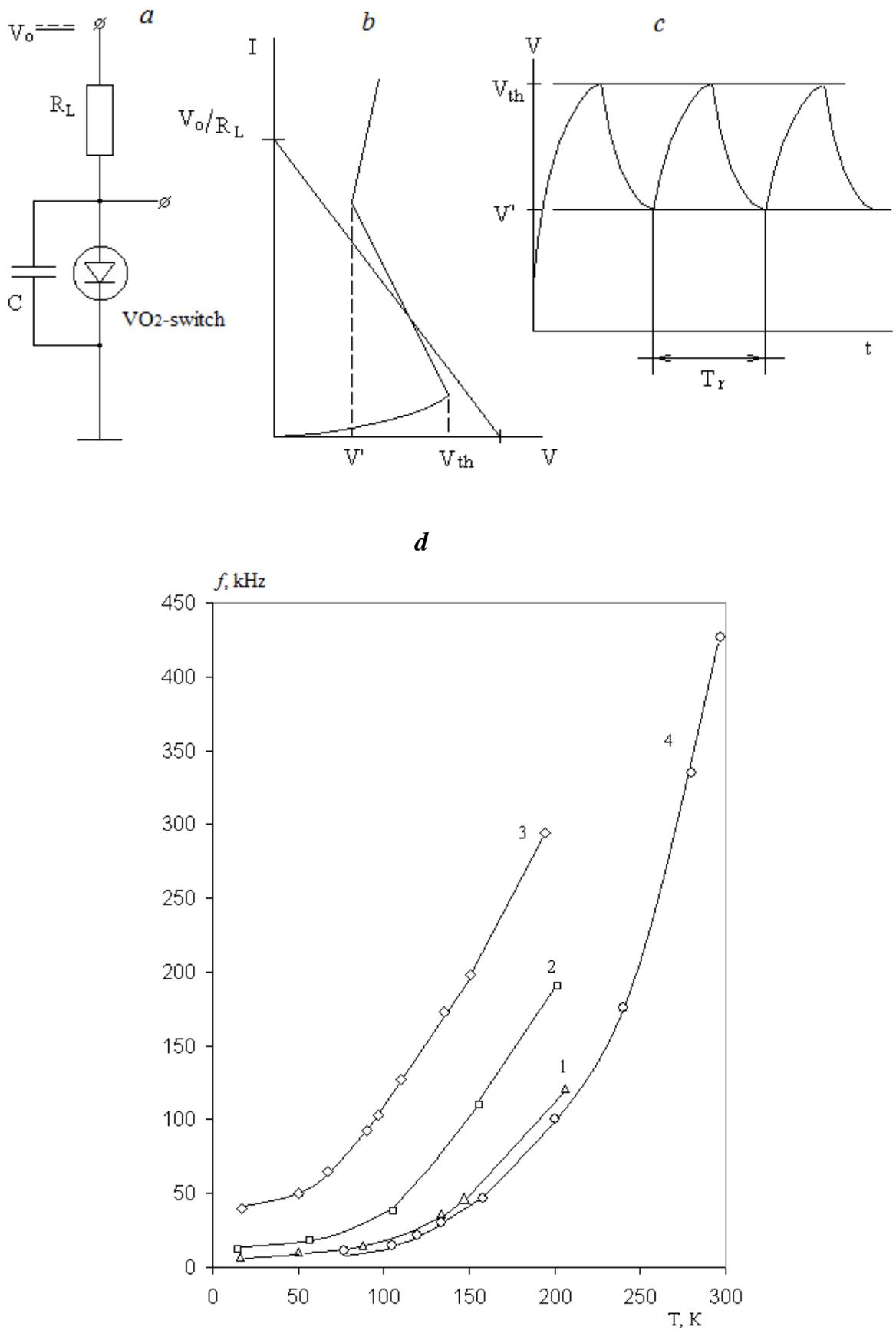

**Fig. 5.** Relaxation oscillations: (a) electrical circuit, (b) schematic $I-V$ curve and load line, and (c) output signal oscillogram. (d) Temperature dependences of $f_r$ for three samples with different threshold voltages (at $T = 100$ K): $V_{th} = 4.4$ V (1, 2), 1.64 B (3) and 7.7 V (4). The values of $V_0$ and $R_L$ are: 80 V and 1.4 M$\Omega$ (1); 95 V and 1 M$\Omega$ (2); 40 V and 0.6 M$\Omega$ (3); 100 V and 1.05 M$\Omega$ (4).

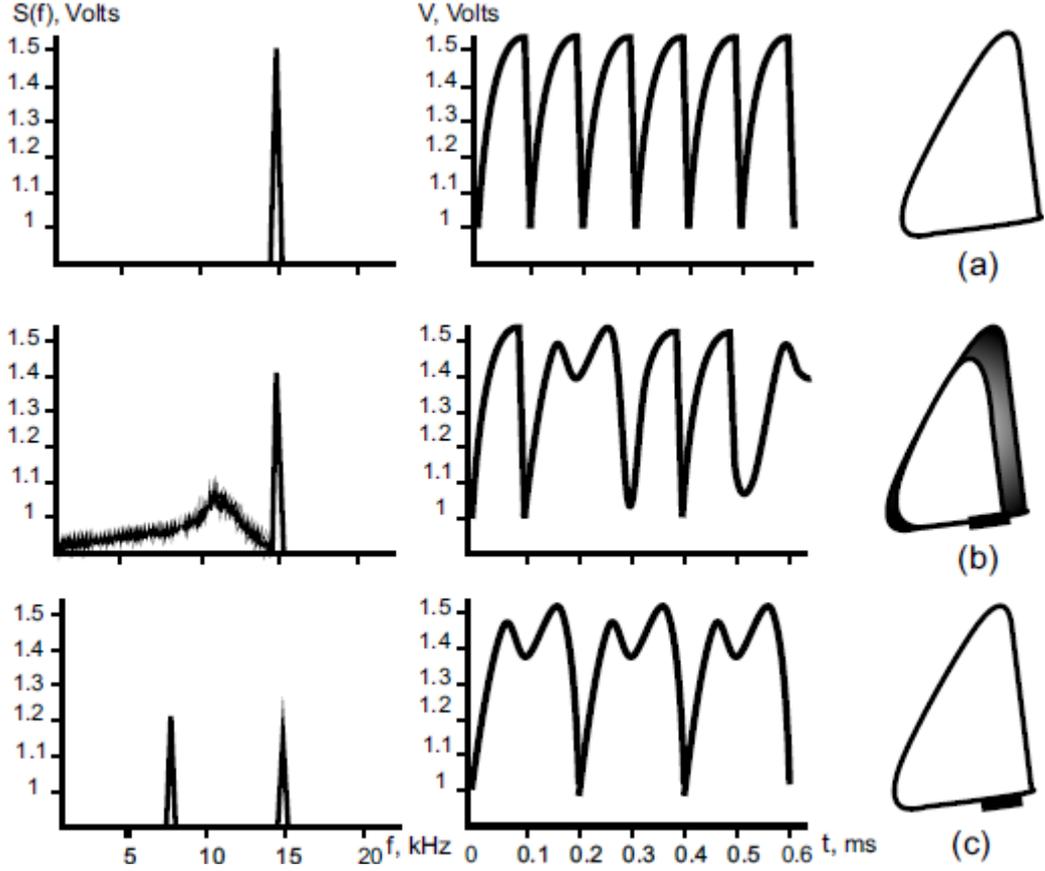

**Fig. 6.** Oscillation spectra, voltage oscillograms, and phase portraits (*I-V* orbits) for different values of *Ω*: (a) *Ω*1, frequency locking; (b) *Ω*2, chaotic oscillations; and (c) *Ω*3, period doubling. *Ω*1 < *Ω*2 < *Ω*3 [36].

Electrical self-oscillations across out-of-plane threshold switches based on VO$_2$ have also been studied in [6]. It is argued there that, beside integration as inverters and oscillators in high-speed circuits, a prospective potential application of these results may be connected with the parallel production of periodic signals in large-scale integrated circuits, analogous to the action potentials in neural architectures. In the next section, we present our results concerning synchronization of a pair of coupled oscillators based on VO$_2$ switches.

### 3. Coupled VO$_2$ oscillators

### 3.1. Background

Nowadays information processing is widely performed by CMOS circuits based on Boolean logics. The amount of devices is continuously increased from 1960s by means of device miniaturization, following the so-called Moore's law [37]. The pace of the miniaturization is set by the ITRS roadmap, which accounts for cost-effectiveness of the R&D efforts required to sustain it. This cost-effectiveness has recently been impacted by fundamental issues such as quantum effects but also by a decrease and a diversification of the demand and by the increase of the cost of the equipment needed for large-scale fabrication of nanoscale devices. Fundamental aspects have been partially and temporally circumvented by the use of new device architectures and new materials [38]. However, the miniaturization pace of CMOS devices has slowed down despite the considerable efforts made by the scientific community and the study of nanoscale emerging devices able to assess the tasks of CMOS transistors has flourished [39].

One of the main hurdle CMOS devices have to tackle to keep on the miniaturisation is power consumption. On the other hand, it appears difficult to fabricate alternative nanoscale device that can compete with CMOS concerning power consumption while showing similar computing efficiency once assembled into a circuit. This calls the paradigm of information processing into question: is it only the CMOS technology that reaches its limit or is it Boolean logics [40]? Dropping Boolean logics for processing information would have a very interesting benefit: one could imagine systems where information is not carried by electric charge, the motion of which is intrinsically dissipative. This has already been proposed for Boolean logics (e.g. with electron spins) but many more information carriers could be envisaged if Boolean logics was not mandatory, in particular analog information carriers could be used. The question to know if analog information carriers are less dissipative than electrical charge is still open but they would enable to drop a more global paradigm which is called Von Neumann computation. It is the base of all actual information processing systems and it schematically consists in performing sequential commands, which is well suited to Boolean logics. It is also known to have a limit coined as "Von Neumann bottleneck" related to the sequential nature. The multi-core trend in microprocessors may indicate that this bottleneck is reached and that it is time to think of intrinsically parallel computation architectures.

This context led some research groups to look into how a brain processes information in order to try to mimic it with electronic circuits. Since there are different kinds of neurons operating with different mechanisms, many bioinspired computation architectures have been proposed [41]. Among them oscillatory neural networks (ONNs) have attracted much attention because the basic bricks are oscillators. An oscillator may be fabricated by many different ways with common electronic devices, most of the time with quite large circuits. However, it has been shown that a simple capacitor made of an insulator featuring a MIT can oscillate between a high-resistance and a low-resistance state when an appropriate bias is applied to it, as is discussed in Section 2.3 above and in the works [42-44]. This means that the basic block of ONNs can be very small, which explains the attractivity of such circuits for large integration.

From a theoretical perspective, ONNs were investigated by Winfree in mid-1960s [45], Kuramoto in mid-1970s [46] and an important mathematical analysis work was done by Crawford in early 1990s [47]. These theoretical contributions are nicely reviewed by Strogatz in ref. [48], however the choice of MIT-based capacitors as basic blocks makes the modelling sensitive because these devices are highly non-linear and typical circuit simulators hardly handle this kind of devices. The question of the behaviour of large arrays of such non-linear devices is still open and may require new kind of simulators, able to account for chaotic behaviour of MIT-based oscillators [36, 49].

Upon this basis, the use of ONNs for information processing was more clearly described by Hoppensteadt and Izhikevich in late 1990s. The idea was to use ONNs to perform associative tasks such as pattern recognition [50]. This is done by capacitively coupling oscillators and checking their synchronisation to assess the proximity of their bias. Indeed synchronisation of oscillators is a crucial feature for the operation of ONNs. It has been widely studied because it is involved in many natural phenomena. It has been particularly suggested that synchronisation was a way for the visual cortex to distinguish an object from the background of an image [51]: the cortical neurons connected to the retinal cells receiving parts of the image that have common properties (shape, colour, movement) are synchronized. Similar operation has also been found in olfactive nervous systems of insects [52]. Mimicking such synchronization-based networks could provide very interesting applications and the most famous theoretical proposition is the associative memory network by Hoppensteadt and Izhikevich [50, 53]. The interest for this network has been renewed by studies dealing with emerging nanodevices, which are able to produce electrical oscillations, so that several attempts of practical implementations of ONNs were proposed [54]. In particular, the recent raise of studies on MIT materials triggered the publication of results on ONNs based on MIT capacitors [55-59].

At present, the exact requirements of practical circuits are being investigated. The external circuitry, which will be required as interface between oscillators and the outer world, has been proposed [60] and the parasitic impact brought about by interconnections on the characteristics of oscillators has been studied [6]. However, other questions still have to be addressed. For instance, the reliability of MIT materials after many oscillations has to be more studied because it may be an issue for applications or at least a source of device-to-device variability, as shown by early papers on this topic [44, 61]. This leads to another question of prime importance: how large can the device-to-device variability while maintaining associative properties? This variability may concern intrinsic frequency and amplitude of oscillations and RLC values of interconnects for instance. It is expected that the tolerance on variability is related to the strength of the coupling between pairs of oscillators, i.e. the value of coupling capacitances in the case of capacitive coupling. The results presented hereafter (and preliminarily reported in [59]) give an overview of the influence of coupling strength on the behaviour of pairs of oscillators.

### 3.2. Experimental results

VO₂ films were obtained by electron-beam evaporation from a vanadium target under oxygen atmosphere on c-cut sapphire substrate and were subsequently integrated in two-terminal planar devices fabricated as described in the work [62]. Such a device can be use either as a fast electrical switch or, within specific excitation conditions, as a relaxation oscillator [6, 63]. Typical images of the realized devices (as basic oscillator blocks) having different dimensions are presented in Fig. 7. The device images of the second row in Fig. 7 show the modification of the VO₂ material color (reflectivity) as an indication of its electrically-induced insulator-to-metal transition while applying on the planar electrodes a voltage having a value of 65 V superior to the threshold voltage initiating the MIT.

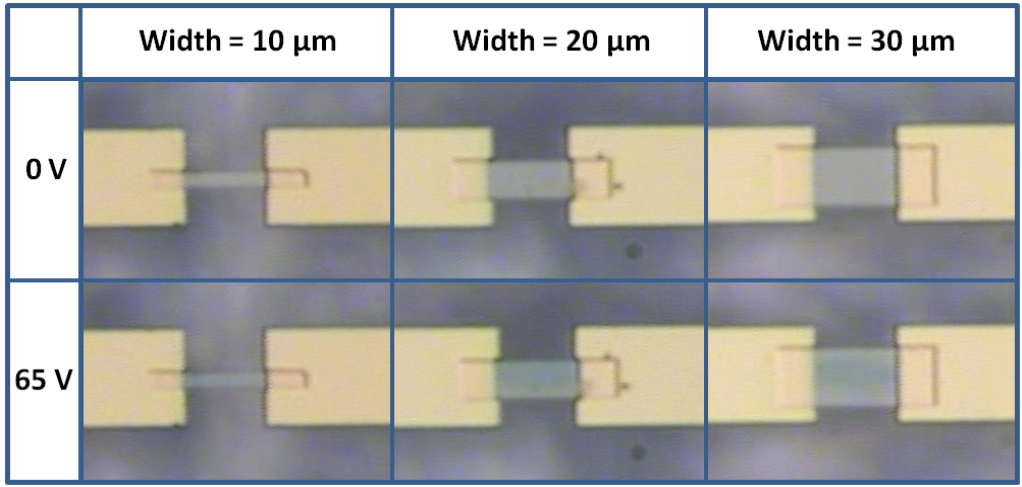

**Fig. 7.** Optical microscopy images of planar devices integrating VO₂ patterns with different widths (30, 20 and 10 μm) and similar length (10 μm) under two different voltages, of 0 V (material in the insulator state) and 65 V (material in the metallic state).

Depending on their geometrical parameters, on the values of the current excitation and device temperature, the frequency of these VO₂-based oscillators can be conveniently and finely adjusted within specific desired values (typically between tens of kHz up to 1 MHz [63]).

Coupling experiments may be carried out very simply by connecting a capacitor between two oscillators (noted as oscillator 1 and 2), as sketched by Fig 8(a). In the measurements presented here, an external variable capacitor (noted as $C_C$) was used. Each oscillator is connected to a series resistor (50 Ω) biased by a carefully chosen current (noted as $I_1$ and $I_2$ for respectively oscillator 1 and 2), in our case a Keithley 2612A current source was used. To determine the bias current, static characteristics of each oscillator is first recorded in $V$-mode, Fig. 8(b), and $I$-mode, Fig. 8(c). The NDR region is observed in $I$-mode characteristics between 0.25 and 2.5 mA on both devices so the current biases have to be chosen in this range to trigger oscillatory behaviors in them. Currents of respectively 1 and 1.5 mA were chosen to bias oscillator 1 and 2, which featured very different oscillation frequencies (respectively at 3.5 and 20 kHz), as shown by Fig. 9(a). However they synchronize when a capacitor with a high enough capacitance ($C_C$=100 nF) is coupling them (Fig 9(c)). The synchrony occurs both in frequency and in phase, as shown by the phase portraits shown in Figs 9(b) and 9(d): without the capacitor, the phase portrait displays disordered points (Fig. 9(b)) whereas with the 100 nF capacitor the points gather around a right segment (Fig. 9(d)), which evidences phase-locking. Actually when the capacitance value is increased up to 100 nF, a progressive frequency locking occurs, characterized by a convergence of the oscillation frequencies of both oscillators. It is observable on oscillograms and phase portraits related to each capacitance value (see Section 5 below for supplementary data) but a convenient way to discern it is to look at the Fourier Transform of the signals as a function of the frequency (Fig. 10), which in this case shows that the final locking frequency is about 4 kHz. The coupling can be even stronger (see supplementary data for $C_C > 100$ nF in Section 5), in this case the phase portraits are more linear but then the applicative sense coupling is lost: oscillators are always synchronised whatever their bias. Neural networks instead expect weak coupling to provide useful applications such as pattern recognition [49, 50], therefore future studies should focus on values of $C_C$ well below 100 nF.

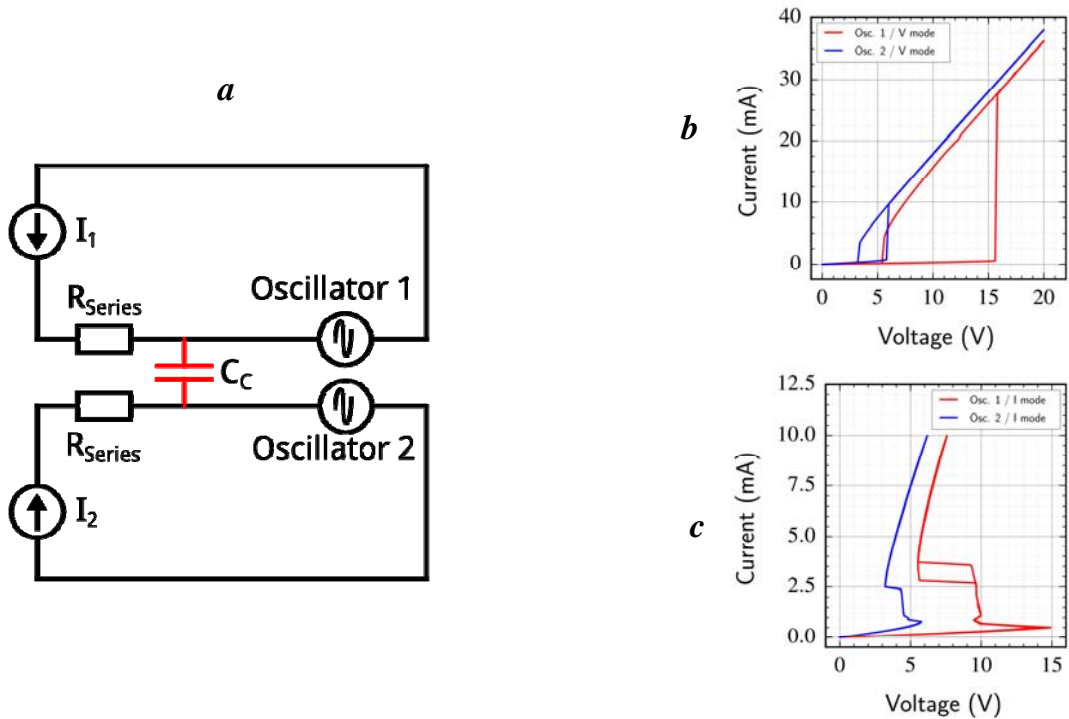

**Fig. 8.** Scheme (a) and *I-V* curves recorded in the sweep voltage (*V*-mode) (b) and sweep current (*I*-mode) (c).

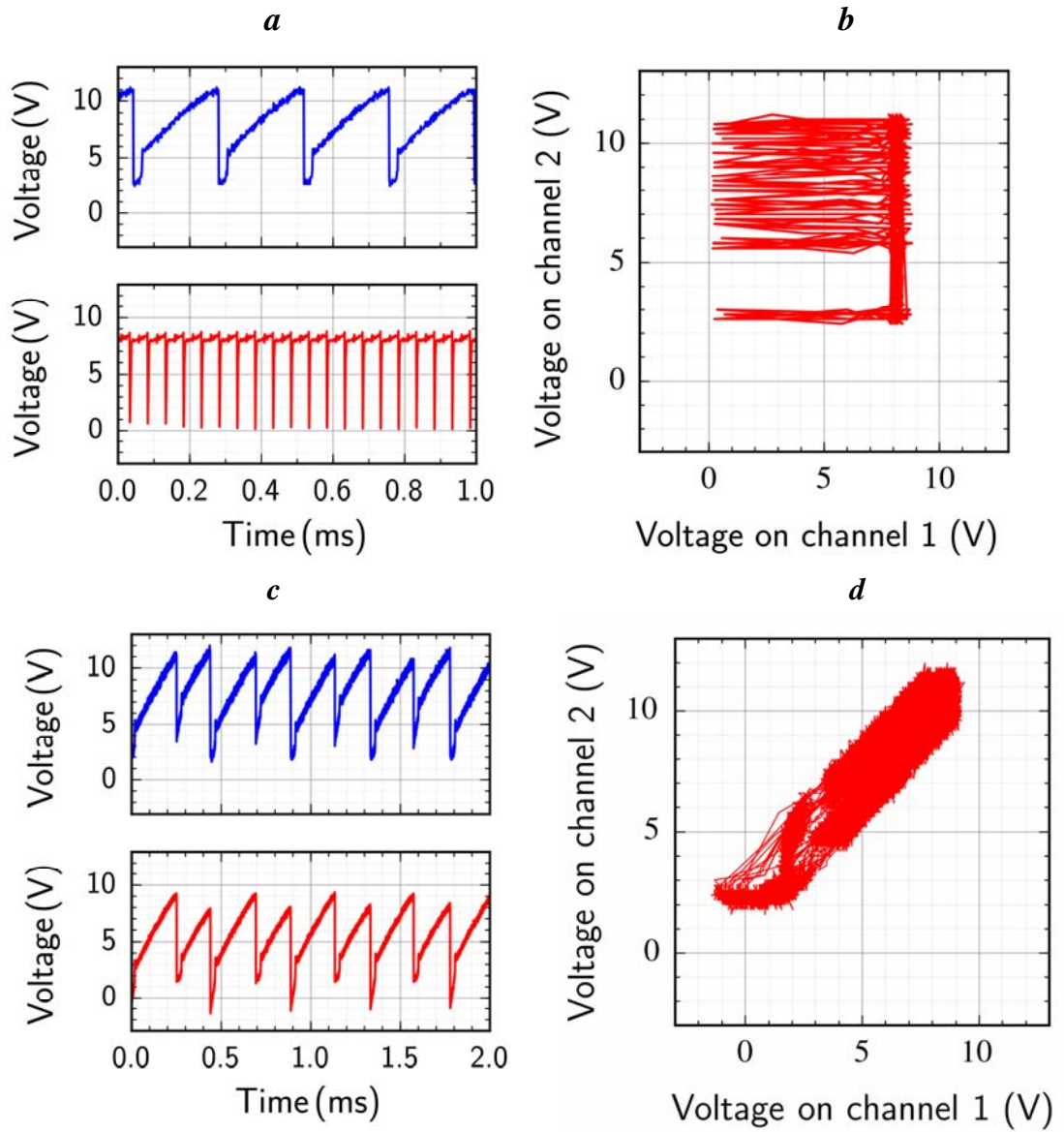

**Fig. 9.** Oscillograms (a, c) and phase portraits (b, d) of two oscillators operating in the uncoupled (a, b) and coupled (c, d) modes. Coupling capacity is $C_C = 100$ nF.

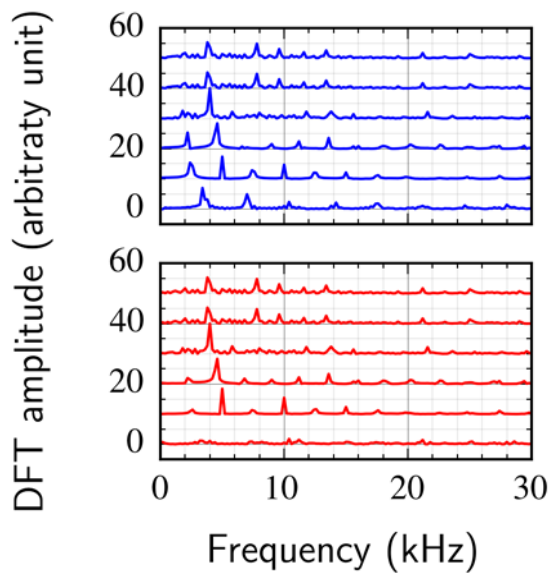

**Fig. 10.** Discrete Fourier Transform (DFT) of oscillations of two devices for six values of $C_C$ (from the bottom up): 10, 50, 100, 500, 1000, and 5000 nF.

## 4. Conclusion and outlook

As is discussed in Section 2, vanadium dioxide undergoes the Mott MIT and the two-terminal MOM devices with $VO_2$ films demonstrate electronic switching associated with this transition. In electrical circuit containing such a switching device, relaxation oscillations are observed if the load line intersects the *I–V* curve at a unique point in the NDR region. This oscillatory behavior is prospective for designing elements of dynamical neural networks based on coupled oscillators.

We emphasize that neurocomputers offer a massively parallel computing paradigm by mimicking the human brain (see discussion in Section 3.1). Their efficient use in statistical information processing has been proposed to overcome critical bottlenecks with traditional computing schemes for applications such as image and speech processing, and associative memory [64]. In neural networks information is generally represented by phase/frequency (ONNs) or amplitude (e.g., cellular neural networks). In contrast to mature amplitude-based neural networks, nascent oscillatory neurocomputing circuits offer more efficient information processing, since they best simulate dynamical binding used in olfactory, auditory and visual systems of the brain [65].

Nonetheless, ONNs have not yet been efficiently realized on the basis of conventional CMOS devices because of large power consumption and high circuit complexity of such CMOS-based implementations which have led researchers to explore the use of emerging technologies for such circuits [64]. Although they provide intriguing properties, previously proposed ONN components based on emerging technologies have not offered a complete and practical solution to efficiently construct an entire system. Fore example, spin-torque oscillators (STOs) coupled with spin diffusion current [66] with electrical read-out has been proposed to construct ONNs [67]. In spite of their novelty and scalability, coupled STOs suffer from high bias currents (in the order of mA) and slower speeds that are limited by the angular spin precession velocity in nano-magnets; in addition, coupling via spin-diffusion current is low power, but localized as the spin diffusion lengths are in the order of μm at room-temperature [57]. An alternative approach is based on synchronized charge oscillations in strongly correlated electron systems [55] such as e.g. vanadium dioxide exhibiting an electronic Mott metal-insulator transition, which is currently considered as one of the most promising materials for neuromorphic oxide electronics [68]. It should be noted however that, alongside vanadium dioxide, other approaches (such as aforementioned STO or resonate body transistor oscillators (RBO) [69]) and other TMOs (for instance, Ta and Ti oxides [70]) have also being developed and considered as prospective candidate techniques for the ONN ideology implementation.

Thus, the material implementation of bio-inspired neural networks is in its infancy. The measurements presented here (Section 3.2) and a few other ones in the literature [7, 55-58] show that early demonstrations of synchronization can be performed with a few MIT materials-based oscillators. Experiments have to be done on larger arrays to assess the feasibility of basic information processing such as associative memorization or saliency recognition. Upon this basis, the impact of weak reliability and large variability should then be studied since neural networks are expected to be more able to deal with such issues than current Von-Neumann information processing architectures.

### 5. Supplement

In Section 3.2, the oscillograms and the phase portraits are shown for uncoupled oscillators and coupling capacitance $C_C$=100 nF. The following figures are data relative to other coupling capacitances: $C_C$=10 nF (Fig.10), 50 nF (Fig.11), 500 nF (Fig.12), 1000 nF (Fig.13) and 5000 nF (Fig.14).

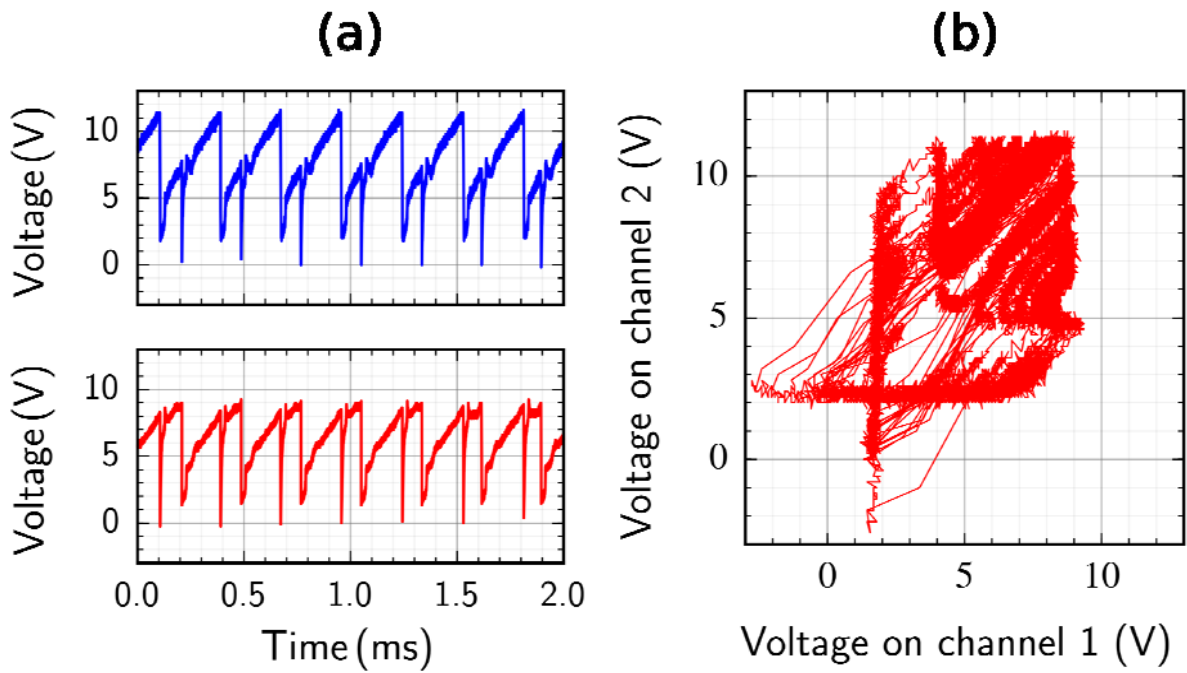

**Fig. 10**. (a) Oscillograms for oscillator 1 (top) and oscillator 2 (bottom) operating in the coupled mode with a coupling capacitance of $C_C$ = 10 nF. (b) Phase portrait related to this coupling.

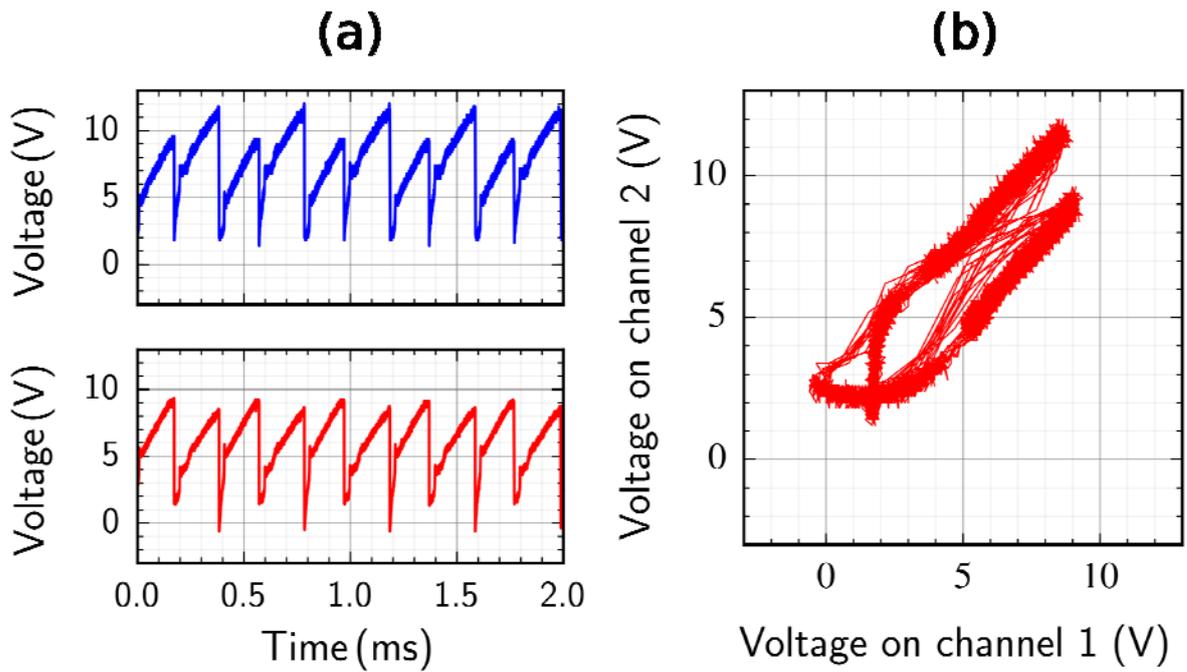

**Fig. 11**. (a) Oscillograms for oscillator 1 (top) and oscillator 2 (bottom) operating in the coupled mode with a coupling capacitance of $C_C = 50$ nF. (b) Phase portrait related to this coupling.

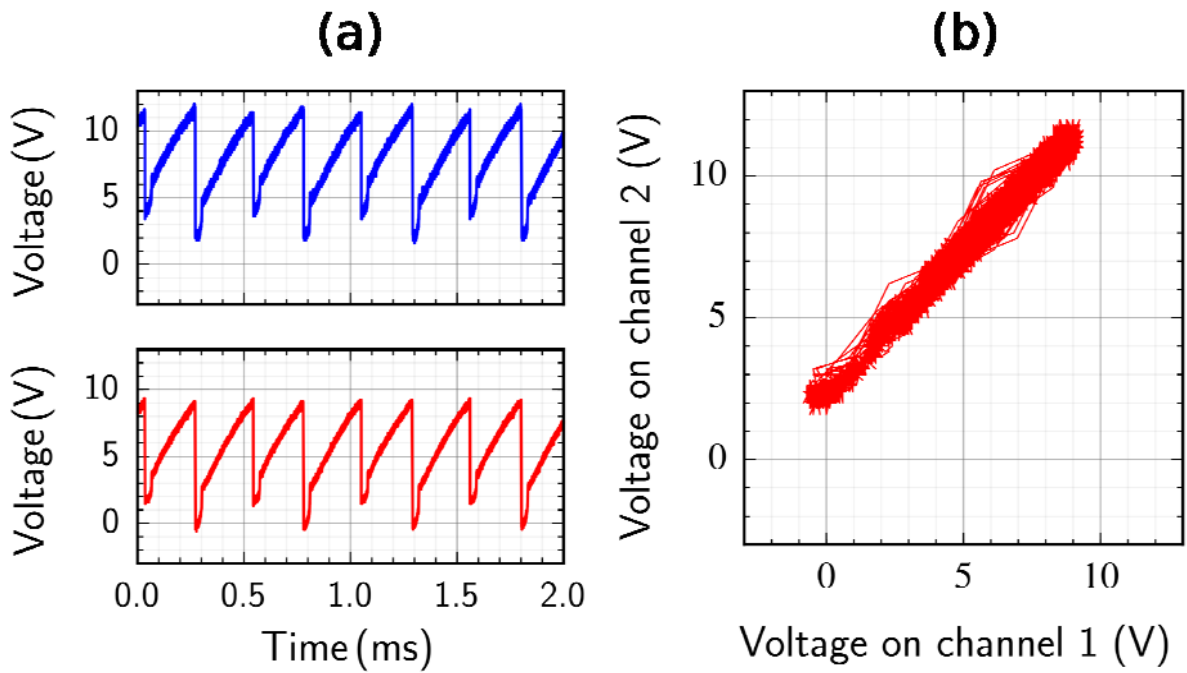

**Fig. 12**. (a) Oscillograms for oscillator 1 (top) and oscillator 2 (bottom) operating in the coupled mode with a coupling capacitance of $C_C = 500$ nF. (b) Phase portrait related to this coupling.

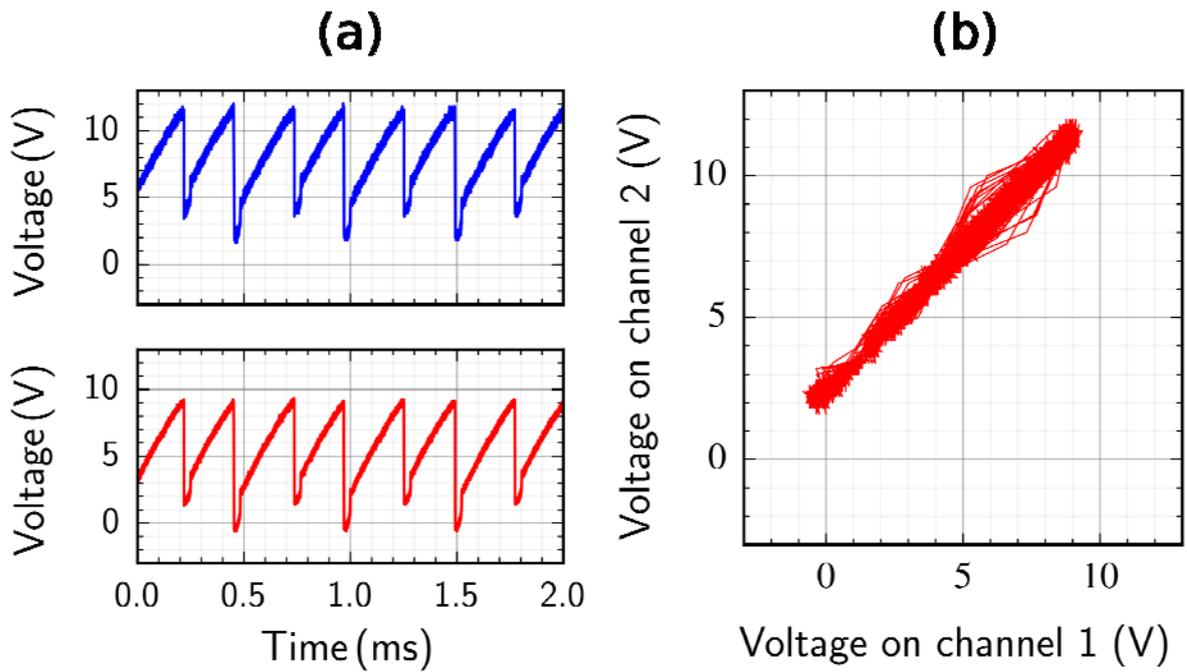

**Fig. 13**. (a) Oscillograms for oscillator 1 (top) and oscillator 2 (bottom) operating in the coupled mode with a coupling capacitance of $C_C$ = 1000 nF. (b) Phase portrait related to this coupling.

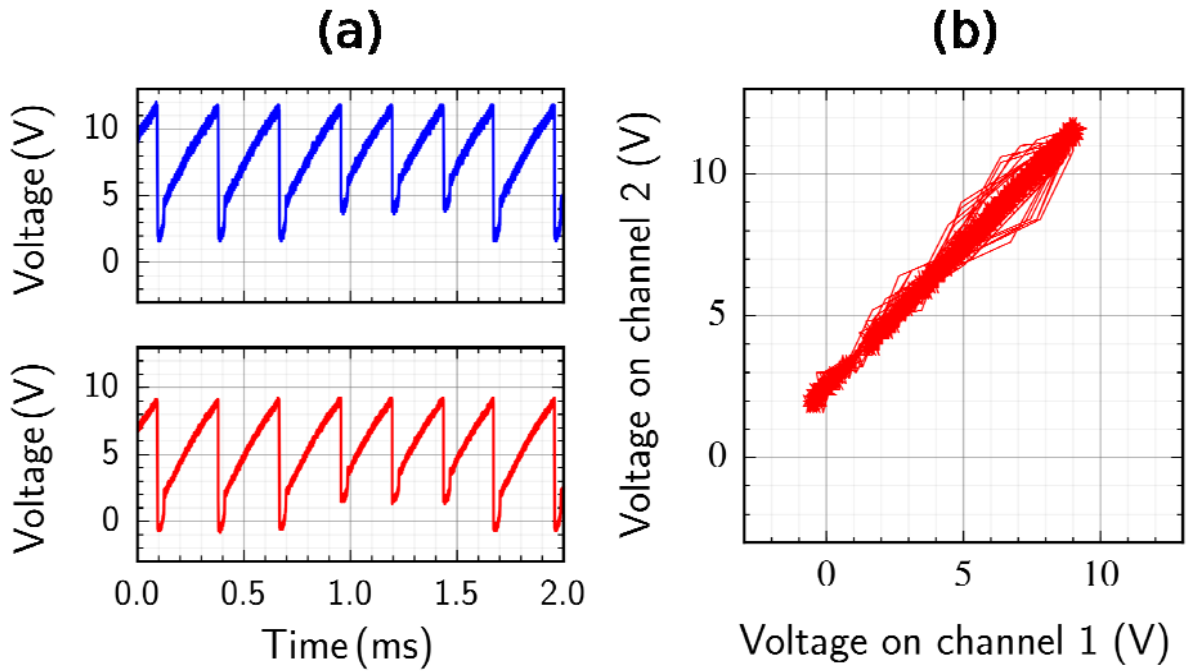

**Fig. 14**. (a) Oscillograms for oscillator 1 (top) and oscillator 2 (bottom) operating in the coupled mode with a coupling capacitance of $C_C$ = 5000 nF. (b) Phase portrait related to this coupling.

**Acknowledgements**

This work was partly supported by the RF Ministry of Education and Science as a base part of state Program no. 2014/154 in the scientific field, Project no. 1426, State Program no. 3.757.2014/K, and RSF grant no. 16-19-00135, (Russia). A. Pergament also acknowledges the financial support of the "Mechnikov Program" (France).